\documentclass[10pt,conference]{IEEEtran}

\usepackage{amsmath,amssymb,amsfonts}
\usepackage{booktabs}
\usepackage{enumitem}
\usepackage{multirow}
\usepackage{float}
\usepackage{graphicx}
\usepackage{epstopdf}
\usepackage{rotating}
\usepackage{xcolor}

\usepackage{subcaption}
\usepackage{graphicx}

\usepackage{graphics}
\usepackage{graphicx}
\usepackage[graphicx]{realboxes}
\usepackage{amsmath,amsfonts,amssymb}
\usepackage{tcolorbox}
\usepackage{multirow}
\usepackage{longtable}
\usepackage{supertabular}
\usepackage{subcaption}
\usepackage{flushend}

\usepackage[linesnumbered,algoruled,boxed,lined]{algorithm2e}
\SetKwRepeat{Do}{do}{while}

\def\BibTeX{{\rm B\kern-.05em{\sc i\kern-.025em b}\kern-.08em
    T\kern-.1667em\lower.7ex\hbox{E}\kern-.125emX}}

\ifCLASSOPTIONcompsoc
  \usepackage[nocompress]{cite}
\else
  \usepackage{cite}
\fi
\ifCLASSINFOpdf
\else
\fi
\begin{document}
%
\title{Automated Misconfiguration Repair of Configurable Cyber-Physical Systems with Search: an Industrial Case Study on Elevator Dispatching Algorithms}
%
%
%
%


\author{\IEEEauthorblockN{Pablo Valle}
\IEEEauthorblockA{
\textit{Mondragon University}\\
Mondragon, Spain\\
pablo.valle@alumni.mondragon.edu}
\and
\IEEEauthorblockN{Aitor Arrieta}
\IEEEauthorblockA{
\textit{Mondragon University}\\
Mondragon, Spain\\
aarrieta@mondragon.edu}
\and
\IEEEauthorblockN{Maite Arratibel}
\IEEEauthorblockA{
\textit{Orona}\\
Hernani, Spain\\
marratibel@orona-group.com}}

%
%

\markboth{IEEE Transactions on Software Engineering,~Vol.~14, No.~8, August~2015}%
{Shell \MakeLowercase{\textit{et al.}}: Bare Demo of IEEEtran.cls for Computer Society Journals}
%



\IEEEtitleabstractindextext{%
\begin{abstract}
Real-world Cyber-Physical Systems (CPSs) are usually configurable. Through parameters, it is possible to configure, select or unselect different system functionalities. While this provides high flexibility, it also becomes a source for failures due to misconfigurations. The large number of parameters these systems have and the long test execution time in this context due to the use of simulation-based testing make the manual repair process a cumbersome activity. Subsequently, in this context, automated repairing methods are paramount. In this paper, we propose an approach to automatically repair CPSs' misconfigurations. Our approach is evaluated with an industrial CPS case study from the elevation domain. Experiments with a real building and data obtained from operation suggests that our approach outperforms a baseline algorithm as well as the state of the practice (i.e., manual repair carried out by domain experts).
\end{abstract}

\begin{IEEEkeywords}
Cyber-Physical Systems, Repair, Debugging, Configurable Systems.
\end{IEEEkeywords}}

\maketitle

\IEEEdisplaynontitleabstractindextext

%
\IEEEpeerreviewmaketitle


\section{Introduction}\label{sec:introduction}





\IEEEPARstart{C}{yber-Physical Systems} combine digital cyber computations with parallel physical processes~\cite{derler2011modeling,baheti2011cyber,alur2015principles}. In such systems, digital technologies, such as computational units, low and high-level software and communication protocols interact among them to control a physical process through sensors and actuators~\cite{derler2011modeling}. 
In practice, most CPSs deal with parameters. For instance, a heavy duty lifting system involved more than 2,000 configuration parameters~\cite{fischer2021testing}. The behavior of CPSs can significantly change depending on these parameters. This often causes misconfigurations, even when selecting parameters that are within the ranges provided by the manufacturer~\cite{han2022control}. A recent study showed that 19.6\% of UAV-specific bugs were caused by parameters~\cite{wang2021exploratory}. Garcia et al.~\cite{garcia2020comprehensive} found that 27.25\% of autonomous vehicle bugs were caused by incorrect configurations. In our industrial case study, which involves the traffic dispatching algorithm of a system of elevators, around 55\% of the issues assigned to the traffic team are solved through configuration changes. Therefore, it is paramount to leverage automated and scalable techniques to automatically repair CPS misconfigurations. However, this involves four core challenges: 

\begin{enumerate}
\item \textbf{Challenge 1 -- Expensive execution of the tests:} It is well-known that executing CPS tests is highly time-consuming~\cite{abdessalem2018testing,abdessalem2018testingVision,menghi2020approximation,menghi2019generating,nejati2019evaluating,haq2022efficient,humeniuk2022search,abdessalem2020automated}. This is because, as the execution of tests is carried out at system level, CPSs involve compute-intensive models to simulate the physical part of the system (e.g., models of electrical engines, dynamics of a system). This makes the computation of the fitness to assess how close the algorithm is from repairing the misconfiguration expensive. For instance, in our industrial case study, executing a test case takes around 5 minutes.
\item \textbf{Challenge 2 -- Large configuration space:} Since configurable CPSs involve many parameters, the amount of possible configurations that a CPS can have is huge. Subsequently, testing all of these configurations is computationally unfeasible~\cite{perrouin2010automated,arrieta2019search,henard2014bypassing,wang2015cost,marijan2013practical,hervieu2016practical}. Furthermore, it is usually unknown which the reason (i.e., the parameters) that causes the misconfiguration is. 
\item \textbf{Challenge 3 -- Multiple requirements:} Multiple failing requirements may exist. Some of them might be independent from one-another~\cite{abdessalem2020automated}, while others may be conflicting (e.g., in our case study, better energy consumption could lead to passengers needing to wait more). Therefore, the repair algorithm shall be approached as a many-objective optimization problem. 

\item \textbf{Challenge 4 -- Prioritize severe failures:} The repair technique needs to resolve failures in their order of severity~\cite{abdessalem2020automated}. For instance a test case that shows a passengers' average waiting time (AWT) of 55 seconds is more critical than one showing 35 seconds. Therefore, similar to other CPS repairing techniques~\cite{abdessalem2020automated}, our algorithm shall give priority to more critical test cases over the less critical ones.


\end{enumerate}

On the one hand, there are approaches that target the problem of repairing misconfigurations~\cite{krishna2020cadet,xiong2014range} of configurable software. However, such approaches only cover the second aforementioned challenge. On the other hand, Swarmbug~\cite{jung2021swarmbug} focuses on repairing misconfigurations of swarm robots, which can be considered CPSs. However, Swarmbug~\cite{jung2021swarmbug} solely focuses on one specific objective (e.g., not crashing), therefore, not tackling the third and fourth challenges that our industrial case study requires.

In this paper we propose an automated repairing approach specifically targeting CPSs' misconfigurations. Specifically, we tackle this by recasting the misconfiguration repair problem to that of a many-objective search problem. To deal with the aforementioned first challenge, we propose an algorithm that follows a single population-based approach. Multiple population-based algorithms, such as genetic algorithms, are not appropriate for this context because the repair process requires interaction with the simulator for executing test cases. Such algorithms require a large population, and the large test execution time would lead the algorithm to require too much time to converge. This could eventually lead to scalability issues in the context of CPSs. To deal with the second challenge, our repairing approach implements a strategy that permits measuring the suspiciousness of each parameter. This permits, as the search process evolves, increasing the probability of selecting suspicious parameters to provide a new patch. As a result, in the beginning of the search, our approach focuses on exploring which the critical parameters can be. As the search evolves, the algorithm starts to focus on the exploitation by targeting suspicious parameters. To deal with the third challenge, our approach includes a Pareto-optimal archive-based strategy to select and evolve potential misconfiguration patches. This permits focusing on more than one requirement at the same time when repairing the misconfiguration. To deal with the last challenge, search objectives are prioritized based on their severity level.



Our main contributions can be highlighted as follows:

\begin{enumerate}
\item  We propose a scalable and automated approach to repair misconfigurations in CPSs.
\item We integrate the approach with an industrial case study from Orona, one of the largest elevator companies in Europe. The case study involves the traffic dispatching algorithm, a highly configurable software system. 
\item We empirically evaluate our approach by using a real scenario in which Orona's engineers had to manually intervene in the misconfiguration repair process. Our repairing technique not only outperforms a baseline algorithm, but also the manually derived repairing patches by Orona's domain experts. 
\item We extract key lessons learned from the application of our approach in an industrial case study, and provide applicability guidelines in order our approach to be adopted by other CPS developers.
\end{enumerate}

The rest of the paper is structured as follows: Section~\ref{sec:casestudies} explains our industrial case study, how the testing is carried out and why misconfigurations occur. In Section~\ref{sec:approach} we present our approach to repair misconfigurations in our industrial context. Section~\ref{sec:eval} presents how we evaluated our approach. We extract key lessons learned and we explain the required changes in our approach to be applied in other CPSs in Section~\ref{sec:lessons}. We position our work with relevant studies in Section~\ref{sec:relatedwork}. We conclude and present future work in Section~\ref{sec:conclusion}.


\section{Industrial Case Study}\label{sec:casestudies}

Our repair algorithm is applied in an industrial case study from the elevation domain. This section explains the different details of the case study.



\begin{figure}[ht]
\centering
\includegraphics[width=0.5\textwidth]{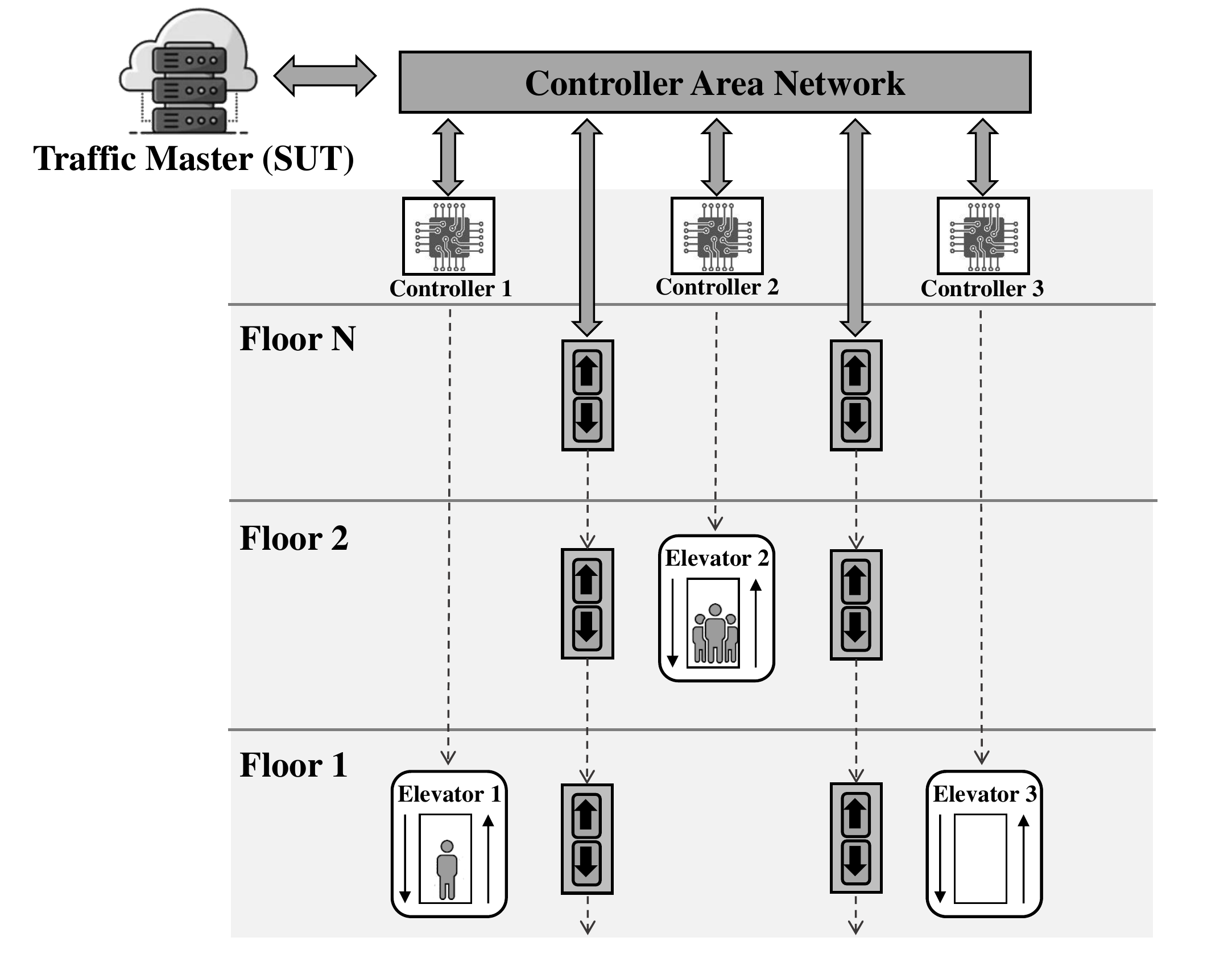}
\caption{Overview of our industrial case study}
\label{fig:casestudy}
\end{figure}

\textbf{The Cyber-Physical System:} Figure \ref{fig:casestudy} shows an overview of the CPS. A system of elevators is a complex CPS, whose goal is to transport passengers from one floor to another safely while trying to provide the highest comfort as possible. In this system, a passenger registers a call in a floor by pushing a call button. This information is transferred to the traffic master through a Controller Area Network (CAN) bus. The traffic master, after collecting other CPS information (e.g., position of each elevator, elevator occupancy), assigns one of the available elevators to each active call. This assignation can be carried out through different objectives (e.g., reducing the passengers' waiting times, reducing energy consumption). When the call is assigned, the elevator attends the passenger.

\textbf{The System Under Test (SUT):} Our SUT is the traffic dispatching algorithm (i.e., dispatcher), which is an important module inside the traffic master. To deal with different functionalities and priorities, the dispatcher is highly configurable through parameters. Different traffic dispatching algorithms exist in Orona, and each of them encompasses one configuration file. The number of potential configurations of each dispatcher is over trillions. 

\textbf{Test Executions:} Three different phases are undertaken when testing the dispatching algorithm~\cite{ayerdi2020towards,gartziandiamachine}: the Software-in-the-Loop (SiL), the Hardware-in-the-Loop (HiL) and Operation. 
Our algorithm is designed for the first phase, i.e., the SiL test level. At this stage, a domain-specific simulator, i.e., Elevate\footnote{https://peters-research.com/index.php/elevate/}, takes as input (1) the dispatching algorithm's executable, (2) the building installation, (3) the configuration file and (4) the passenger file. The passenger file is considered the test input, and it involves a set of passengers traveling through different floors in a building. Each passenger has different attributes, such as, its arrival time (i.e., time at which the passenger arrives to the floor and pushes the button), arrival floor (i.e., floor at which the passenger arrives), destination floor (i.e., floor at which the passenger is traveling to), passenger weight, etc. When a test is executed, Elevate returns a file with the results of the simulation (e.g., waiting time required by each passenger, their traveling time, energy consumption, distance traveled by each elevator). This information is parsed and the necessary test oracles are employed to assess the quality of the execution of the test.



\textbf{Functional performance requirements:} When executing test cases, besides considering certain functional requirements, we focus on \textit{``functional performance requirements''}. Functional performance is defined as \textit{``the properties derived indirectly from the output of the system, rather than the system's efficient usage of the computational resources''}~\cite{gartziandiamachine}. These properties are directly employed for evaluating the functional performance requirements of Orona's dispatching algorithms. The properties involve metrics from the elevator traffic domain, such as the Average Waiting Time (AWT) of passengers, the Average Transit Time (ATT) of passengers, Longest Waiting Time (LWT), Longest Transit Time (LTT), number of engine starts, traveled distance by each elevator or consumed energy. Note that configuration changes affect functional performance requirements, whereas functional requirements (e.g., ensuring that reverse journeys do not take place) are, in principle, not affected by such changes. 


\textbf{Why misconfigurations occur and how they are handled:} The dispatcher has different parameters to accommodate different functionalities that have a direct impact on the CPS performance. However, it is noteworthy that a configuration may perform well in one installation of elevators, while not well in another one, causing a misconfiguration. This is because the performance of a system of elevators largely depends on (1) the type of building and its composition and (2) how its traffic flow is. Regarding the former, the performance can vary depending on aspects like number of elevators in a building, the number of floors the building has, whether all elevators attend all floors or not, etc. For some types of buildings, some configurations are more appropriate than others. As for the latter, the traffic is also different depending on the type of buildings. For instance, the traffic flow is completely different in a hospital and in a residential building. While in a hospital inter-floor travels are common, in a residential building most of the calls are from the base floor to the floor where the apartment is and vice-versa. When a system of elevators shows a poor performance, its traffic flow is reproduced at the SiL test level to debug and try to improve its performance through changing parameters. If a new set of parameters improves the system performance, then, the original configuration is considered a misconfiguration. It is important to note that in our industrial case study, a misconfiguration might not be detected nor foreseen before the system is in operation due to the CPS exposition to uncertainty \cite{han2022uncertainty,han2022elevator}.






\section{CPS Misconfiguration Repair Method}\label{sec:method}
\label{sec:approach}

Algorithm \ref{alg:repair} shows an overview of our repairing algorithm. The algorithm takes as input (1) a faulty configuration file $C$, composed of $N$ number of parameters, i.e., $C=\{p_1, p_2, ..., p_N\}$; and (2) a test suite, composed of $M$ failing test cases, i.e., $TS=\{tc_1, tc_2,..., tc_M\}$. The first step of the algorithm consists on assessing the failing configuration file, where all the parameter values are parsed (Line 1) and all test cases are executed (Line 2). When the failing test suite is executed, the values returned by the oracle are used to initialize the Archive (Line 3) and the suspiciousness scores of parameters initialized (Line 4). After that, the algorithm enters into a while loop (Lines 5-11) that ends when the termination criteria are met. These criteria involve (1) fixing the misconfiguration or (2) exceeding the running time.

\begin{algorithm}[ht]
\caption{Overview of our search-based repairing algorithm}\label{alg:repair}
    \KwIn{C //\textit{Faulty Configuration file} \\
        TS //\textit{Test Suite} }
    \KwOut{Archive //\textit{Archive containing improved configurations} \\}
    
    $Patch_0$ $\leftarrow$ getValues(C); \\
    InitialScore$\leftarrow$ executeTestSuite($Patch_0$, TS);\\
    Archive $\leftarrow$ saveToArchive($Patch_0$, InitialScore); \\
    Susp $\leftarrow$ initSusp();\\
    \While {terminationCriteriaNotMet}{ 
        Parent $\leftarrow$ selectAParentArchive(Archive);\\
        $Patch_1$ $\leftarrow$ generatePatch(Parent,Susp); \\
        Score $\leftarrow$ executeTestSuite($Patch_1$, TS);\\
        Susp $\leftarrow$ updateSusp($Patch_1$, Parent, Score, ScoreParent);\\
        Archive $\leftarrow$ saveToArchive($Patch_1$, Score);\\
    }
    \Return Archive;
\end{algorithm}

Inside this while loop, the first step consists in selecting a solution from the Archive (Line 6), which will be the parent. The solution is selected pure randomly. With the selected solution, a potential patch is proposed (Line 7), which consists of changing one or more parameters from the parent solution (Section~\ref{sec:patchgen}). This patch is assessed by executing the failing test suite (Line 8), and the test execution results are obtained and stored as Scores (Section~\ref{sec:testexec}). In a fourth step, the suspiciousness score of each parameter is recalculated (Line 9, Section~\ref{sec:suspicmeas}). Lastly, the Archive is updated (Line 10, Section~\ref{sec:archive}).

\subsection{Patch generation}
\label{sec:patchgen}

A patch in our context refers to a mutation of at least one parameter. Algorithm~\ref{alg:GeneratePatch2b} shows our algorithm for proposing a potential patch. As input, it receives (1) a parent configuration, which corresponds to one configuration in the archive of the algorithm and (2) the suspiciousness ranking of all parameters. First, a parameter to be mutated is selected (Line 4) based on the suspiciousness of each parameter (see Section ~\ref{sec:suspicmeas} for more details on how to compute the suspiciousness score). The higher the suspiciousness, the higher the probability of being selected. The parameter to be mutated is obtained by employing Algorithm~\ref{alg:selectSuspiciousParam}. The selected parameter is mutated (Line 5) by giving a random value within its ranges. After this, it is decided whether a new parameter is mutated (Line 8). The probability of mutating a new parameter decreases as the number of mutated parameters in the new patch increases. We ensure that one parameter is not mutated more than once.

\begin{algorithm}[ht]
\caption{Patch generation algorithm}\label{alg:GeneratePatch2b}
    \KwIn{Parent //\textit{Faulty Configuration} \\
    SuspRanking //\textit{Suspiciousness Ranking}}
    \KwOut{Patch //\textit{Mutated Configuration} \\}

    numOfMutParams $\leftarrow$ 0;\\
    Patch $\leftarrow$ Parent;\\
    \Do{$p < 0.5^{numOfMutatedParams}$}{
      varToMutate $\leftarrow$ selectParam(SuspRanking);\\
        Patch $\leftarrow$ mutate(Patch,varToMutate);\\
        numOfMutParams $\leftarrow$ numOfMutParams +1; \\
        p $\leftarrow$ rand(); //\textit{returns random value 0 to 1}\\ 
    }
    
    \Return Patch;
\end{algorithm}

\begin{algorithm}[ht]
\caption{Suspiciousness-based parameter selection algorithm}\label{alg:selectSuspiciousParam}
    \KwIn{SuspScore = $\{ss_1, ss_2, ..., ss_N\}$ }
    \KwOut{selected //\textit{Index of the selected parameter} \\}
    
    total $\leftarrow$ $\sum_{i=1}^{N}(ss_i);$\\
    iterativeSum$\leftarrow$0; \\
    prob $\leftarrow$ []; \\
    \For{i $\leftarrow$ 1 \KwTo nPop}{
        prob[i] $\leftarrow$ iterativeSum + SuspScore[i]/total; \\
        iterativeSum$\leftarrow$prob[i];\\ 
    }
    prob$\leftarrow$orderAscending(prob);\\
    r$\leftarrow$rand();//\textit{Returns random number 0 to 1}\\
    j$\leftarrow$0;\\
    selected=N;\\
    \While {j$<$N and selected==N}{ 
       \uIf{r$<$prob[j]}{
        selected$\leftarrow$j;\\
       }
       j$\leftarrow$j+1;\\
    }
    \Return parameter(selected); //\textit{translates index of selected to parameter ID}
\end{algorithm}

\subsection{Test suite execution}
\label{sec:testexec}

After the patch is generated, this needs to be assessed. We assess each patch by re-executing  all test cases in the test suite that have failed. We do not execute the passing test cases because executing such test cases would significantly increase the computational time of our approach. Furthermore, for the sake of increasing the efficiency of our repair algorithm, the process of executing test cases is parallelized. When executing the test suite, test oracles assess the performance of the system. In our context, similar to other approaches~\cite{abdessalem2018testing,abdessalem2018testingVision,abdessalem2020automated,menghi2019generating,menghi2020approximation,humeniuk2022search,arrieta2022automating}, test oracles not only provide a boolean verdict (i.e., \textit{Pass} or \textit{Fail}), but also a confidence value. The lower the value, the lower the performance of the CPS in terms of the assessed property by such test oracle.

These oracles' confidence values are used as search objectives to guide the repair algorithm towards finding effective patches. For repairing a CPS, a total of $k$ test oracles may exist. Each of these $k$ oracles acts as an individual objective function in the repair algorithm. For each test case ($tc$) in the failing test suite ($TS$), each of these $k$ oracles returns its confidence value, i.e., $Conf(tc, o_i) \in [-1,0]$, where $o_i$ is the $i$-th oracle. -1 means that the severity of the failure is the highest contemplated one, whereas 0 means that the oracle has passed. The repair algorithm aims at maximizing that confidence value. Therefore, after executing all test cases in $TS$, similar to Abdessalemm et al.,~\cite{abdessalem2020automated}, we obtain the minimum value for each of the test oracles (i.e., the most severe value), converting the repair problem in a many-objective optimization problem that gives priority to the most severe failures, such that:

\begin{equation}
 \label{eq:Objectives}
    \left\{
       \begin{array}{ll}
	 \max Oracle_1(Patch) = \min\limits_{tc \in TS } \{Conf(tc, o_1)\}  \\
	 ...\\
	 \max Oracle_k(Patch) = \min\limits_{tc \in TS } \{Conf(tc, o_k)\}
       \end{array}
     \right.
\end{equation}




As previously explained, executing a test in the context of CPSs is time consuming. Previous studies using compute-intensive CPSs have leveraged surrogate models to accelerate the generation of test cases~\cite{abdessalem2018testing,haq2022efficient,humeniuk2022search,menghi2020approximation}. That is, after a set of test executions, a model is trained with test results, and this model is employed as a substitute of the simulation-based test execution. This permits accelerating the generation of test cases. While we considered to use surrogate models to accelerate the repair process, we noticed that too many simulations were required to obtain a reliable surrogate model. Unlike previous approaches~\cite{abdessalem2018testing,haq2022efficient,humeniuk2022search,menghi2020approximation}, which only use the dimension of the test input, configurable CPSs also need to consider the dimension of parameters, which makes it harder to train a surrogate model. After carrying out a preliminary evaluation with our industrial case study, we noticed that the required time to obtain data for building a reliable surrogate model was similar or even higher than the time required by our repair algorithm to converge. Therefore, the option of using a surrogate model to accelerate the repair process was discarded.

\subsection{Measuring parameter suspiciousness}
\label{sec:suspicmeas}

Based on analyzing the behavior of our industrial case study, and by interviewing domain experts, we noticed that some parameters have a higher influence than others on the system performance. Therefore, we implement a mechanism to measure the suspiciousness of each parameter in $C$. 
The suspiciousness provides a score between 0 and 1, where the higher the suspiciousness, the higher the likelihood of the parameter having an influence in the system performance. The ultimate goal of this strategy is to give a higher probability of being mutated to those parameters having an influence in the system performance.


All configurable parameters start with the same suspiciousness score, which is 0.5. This suspiciousness remains unchanged until the parameter is mutated by the Patch generation algorithm for $N_{susp}$ times (we employed $N_{susp}$ = 5 in our experiments). This permits the algorithm to focus on the exploration phase at the beginning of the search process, while focusing on the exploitation as the search process evolves. Every time a parameter is mutated by the Patch generation algorithm, after assessing the patch, we extract whether the parameter had (1) a positive impact on the performance of the system, (2) a negative impact on the performance of the system or (3) no impact at all. A positive impact of a parameter $p_i$ is considered when the patch is non-dominated by any other patch in the system based on the test results. A negative impact of a parameter $p_i$ is considered when the patch is dominated by the solutions in the archive (i.e., including its original parent). The patch does not have any impact for a parameter $p_i$ when the result of the test shows the same performance as its original parent. After a parameter $p_i$ is selected $N_{susp}$ times, its suspiciousness starts to be computed as follows:

\begin{equation}
    susp(p_i) = \dfrac{P_{p_i}+N_{p_i}}{P_{p_i}+N_{p_i}+S_{p_i}}
\end{equation}

\noindent where $P_{p_i}$ is the number of times that parameter $p_i$ had a positive impact, $N_{p_i}$ is the number of times that the parameter $p_i$ had a negative impact and $S_{p_i}$ is the number of times that the parameter $p_i$ had no impact at all.

Notice that either the positive or the negative impact increase the suspiciousness of a particular parameter. This is because the patch is proposed by mutating the value of a parameter by another random value within its ranges. Therefore, another value in a parameter that previously had a negative impact may have a positive impact on the CPS performance.

Based on our analysis, the suspiciousness of the parameters in the context of CPSs is, in principle, unknown, even with domain expertise. This is, to a large extent, because CPSs highly depend on the context at which they operate. For instance, in the case of our industrial case study, a parameter can have a large impact on the performance of the CPS depending on the type of building (e.g., parameters may behave differently in a residential building with 2 elevators or in a hospital building with 4 elevators). For this reason, we assume there is no prior knowledge of the impact a parameter may have in the context of a CPS. However, our approach for measuring the suspiciousness of parameters can easily be extended to other strategies (e.g., providing the algorithm with an initial suspiciousness score for each of the parameters in the configuration).

\subsection{Updating the Archive}

\label{sec:archive}

Our algorithm uses an archive encompassing non-dominated solutions that are generated by including patches. The first configuration file being updated in the archive is the misconfiguration provoking the failure. After assessing a patch ($Patch_1$) by executing the failing test suite, the archive needs to be updated. Such patch is compared with the rest of solutions in the archive. The comparison is based on the notion of dominance, and similar to other studies~\cite{abdessalem2020automated}, the archive is updated as follows:

\begin{enumerate}
    \item If $Patch_1$ dominates at least one solution in the archive, $Patch_1$ is included in the archive, and the dominated solutions are removed.
    \item If no element in the archive dominates $Patch_1$, but $Patch_1$ is neither dominated by any solution in the archive, $Patch_1$ is included in the archive.
    \item The archive remains unchanged if $Patch_1$ is dominated by at least one solution in the archive.
\end{enumerate}

By following this strategy, there is some risk that the archive increases in size. This would lead the algorithm to need much more time to converge. To overcome this problem, if the archive exceeds certain size, we remove solutions from it. Same as Abdessalem et al.,~\cite{abdessalem2020automated}, the maximum size of our archive is limited to $2 \times k $, $k$ being the number of oracles. However, unlike~\cite{abdessalem2020automated}, instead of randomly removing the solution from the archive, we removed the solution which had the longest Average Waiting Time (AWT). This decision was taken because in the elevation domain, this is the main metric used to assess the performance of a dispatching algorithm~\cite{barney2015elevator}. If two or more solutions encompassed the same highest AWT, the choice is random among those two solutions.


\subsection{Stopping criteria}

The repair process stops given two criteria: (1) all test cases in $TS$ pass, i.e., all oracles in all test cases return the $0$ value; or (2) the search budget is exceeded (i.e., repairing time was exceeded). If the latter happens, it might be the case where the test cases are too demanding. Therefore, the repair process would be converted into a parameter optimization problem. For instance, by analyzing our industrial case study with the elevator dispatching algorithm, we noticed that some test inputs may encompass too many passenger calls in a short time window. In such cases, the CPS may enter in a saturation state, where the only solution would be to include additional elevators to better attend calls, something that is out of the scope of the dispatching algorithm's competence.

\subsection{Decision maker}

When the repairing algorithm stops due to the search budget being exceeded, there might be a high probability that more than one solution exists in the Archive. In such a case, a decision maker (DM) with certain rules would need to select one of the solutions and propose it as a patch. This decision maker is, in our case, domain-specific. The DM was a rule-based algorithm that was designed by involving domain experts in the process. The specified thresholds are configurable because some thresholds may be valid in certain buildings but not in others. The algorithm follows the next procedure to decide which patch to propose:

\begin{enumerate}
\item It first selects all patches where the AWT is less than 25 seconds. This is the threshold that an international standard considers as a good performance of a system of elevators~\cite{cibse2010transportation}. Since the AWT is the most widely employed metric to assess the quality of a system of elevators~\cite{barney2015elevator}, we gave first priority to this metric. If there is no solution meeting that requirement, we select the patch that exhibits the lowest AWT. 
\item If more than one patch remains, the DM prioritizes patches whose test execution showed a lower number of passengers waiting above 55 seconds. That threshold is specified to be below 10\%, which was considered an affordable number. Domain experts considered that waiting nearly a minute is an anti-pattern, therefore, they decided to give priority to those solutions that exhibited a low number of passengers waiting more than 55 seconds.
\item In a third stage, if more than one patch exists in the set of candidate solutions, priority is given to the ATT metric. The DM selects those solutions that have a lower ATT than 45 seconds. If there are no solution meeting that requirement, we select the patch that exhibits the lowest ATT.
\item If multiple patch candidates keep existing, the DM selects those solution whose test execution showed a lower number of passengers having a transit time above 70 seconds. That threshold was specified to be below 10\%, as it was considered an affordable number.
\item After that, in the event that more than one candidate patch existed, the DM selected the patch with lowest LWT, which was considered of higher importance than the LTT. If more than a patch existed, the patch with lowest LTT was chosen. Although the possibilities are remote, it is still possible to have more than one solution. In such a case, the similarity of the configuration files of the candidate patches is compared with the original configuration file through the well-known hamming distance metric. The one which has more similarity is chosen. The reasons are two-fold. On the one hand, engineers are not usually eager to change too many parameters from the original configuration file. This is because, what it is good for certain passenger flows, it may not be good for others. On the other hand, we conjecture that the higher the number of parameters that have been changed, the higher the probability that the solution is overfitted to the failing test suite. Therefore, by means of this mechanism, we aim at reducing the probability for our plausible patch to be overfitted.
\end{enumerate}






\subsection{Patch confirmation}

Since we only use a failing test suite to repair the misconfiguration, the patch needs to be retested. This way, we ensure that the patch is not overfitted to the failing test suite, which is a core problem of automated program repair~\cite{goues2019automated,martinez2017automatic,nilizadeh2021exploring,smith2015cure}. This can be carried out following any kind of state-of-the-art technique. In our case, we use a regression test oracle~\cite{gartziandiamachine} and execute synthetic test inputs (i.e., test inputs based on templates for full-day theoretical passenger profiles~\cite{siikonen2001passenger} and up and down-peak profiles suggested by international elevator standards~\cite{cibse2010transportation}). We ensure that the new patch does not perform worse than the original patch. Besides, we test its functionality by employing metamorphic testing with shorter test cases, as proposed by Ayerdi et al.~\cite{ayerdi2020qos,ayerdi2022performance}.


\section{Evaluation}\label{sec:eval}

 
In our evaluation, we aimed at answering the following two research questions (RQs):

\begin{itemize}
\item \textit{\textbf{RQ1 -- Sanity check:} How does our approach compare to the baseline?} To assess whether the problem to solve is trivial, the first RQ is a sanity check. To do so, we implemented an unguided version of our repairing algorithm. 
\item \textit{\textbf{RQ2 -- Comparison with state of the practice:} How does our approach compare to manual repair carried out by domain experts?} The current practice at Orona is to manually repair the misconfigurations. This RQ aims at comparing whether our algorithm is competent when compared to a manual repair process carried out by domain experts in the company.
\end{itemize}

\subsection{Experimental Setup}

\subsubsection{System Under Test and Building}
We used Orona's Conventional Group Control (CGC) traffic dispatching algorithm~\cite{barney2015elevator}, which has also been used in other studies~\cite{ayerdi2020qos,ayerdi2021generating,ayerdi2022performance,han2022elevator,han2022uncertainty}. Furthermore, we used a real installation to assess our approach. The installation involved a total of three elevators and 12 floors. We used this installation because it was a real case where Orona had to manually intervene to resolve the misconfiguration. Furthermore, the manual misconfiguration process taken by the engineers was well documented. In addition, we also had access to the operational data obtained from the conflicting installation to be used as failing test inputs. In total, we used three failure-inducing test inputs, involving 16 hours of passenger flow each, and between 3,105 and 3,769 passengers in total.

The version of the algorithm we used involved a total of 43 parameters. The total number of potential configurations ascends to over $9.3\times10^{92}$, which makes the search space too large to employ brute force.



\subsubsection{Test oracles}
By carefully analyzing the internal document Orona used to give solution to the conflicting installation, we defined six oracles based on the metrics they were aiming to optimize. Below we explain the selected functional performance metrics:

\begin{itemize}
    \item Average Waiting Time (AWT): It measures the average waiting time of all passengers. The waiting time refers to the time since a passenger registers a call until an elevator arrives to attend her.
    \item Longest Waiting Time (LWT): It measures the longest waiting time experienced by the passengers.
    \item \% of passengers with Waiting Time (WT) above 55 seconds: It measures the percentage of passengers who had to wait more than 55 seconds.
    \item Average Transit Time (ATT): It measures the average transit time of all passengers. The transit time refers to the time since a passenger enters a lift until it arrives to its destination.
    \item Longest Transit Time (LTT): It measures the longest transit time of all passengers.
    \item \% of passengers with Transit Time (TT) above 70 seconds: It measures the percentage of passengers who had a transit time above 70 seconds.
\end{itemize}


When repairing this misconfiguration, the domain experts aimed at improving as much as possible the functional performance metrics listed above. Therefore, in the context of this study, we opted for being aggressive with the thresholds. Therefore, all thresholds were set to 0. We acknowledge that these values are unfeasible to obtain. However, this way the comparison with the manual approach is fairer. Furthermore, we also wanted to assess the patch that the DM selected.


\subsubsection{Execution platform}

Elevate version 8.19 was used as simulator for executing the tests. The experiments were carried out using a PC with a Windows 10 operating system, with a CPU Intel Core i5 7th generation, and a 16 Gb RAM. 


\subsubsection{Baseline algorithm and state of the practice comparison}


As baseline algorithm, we developed an unguided version of our repairing algorithm. Two core differences exists between the unguided version and the repair algorithm proposed in this paper: (1) the unguided version saves all configurations in the archive and (2) the parameters to be mutated are considered all to have the same suspiciousness score (i.e., the suspiciousness is not measured in this version). It is noteworthy that this baseline is stronger than a pure Random Search (RS), which is the usual baseline algorithm used to assess search-based software engineering problems~\cite{wang2015cost,arrieta2019pareto,arrieta2019search,arrieta2017employing,mcminn2011search,di2007search}. This is because, RS would take the initial failing configuration and propose some patches based on our patch generation approach (Algorithm~\ref{alg:GeneratePatch2b}). However, with RS, these generated patches would not evolve anymore. Conversely, with our unguided approach, we give the option of evolving patches in the archive, leading to higher probabilities of finding a patch.

As for the comparison with the state of the practice, for the building installation used, we had data from engineers from Orona. Specifically, when the issue was raised, engineers from Orona proposed different potential patches (i.e., different configurations of the dispatcher). We compared the results obtained by our algorithm with the patches proposed by the domain experts. Six different patches were provided by Orona's engineers.

\subsubsection{Evaluation Metrics}

As our algorithm is Pareto-compliant, we had to assess all the solutions in the archive as a whole. Because of this, and based on related guidelines~\cite{wang2016practical,li2022evaluate}, we used the \textit{Hypervolume (HV)} quality indicator. The \textit{HV} is one of the most widely employed metrics to assess Pareto-based search algorithms~\cite{wang2016practical,li2022evaluate,shang2020survey}. The \textit{HV} measures the volume in the objective space of a search algorithm, and has many advantages~\cite{shang2020survey}, such as, (1) being Pareto compliant, (2) being able to evaluate convergence and the diversity of a solution set simultaneously and (3) only requiring one reference point.

Besides the \textit{HV} quality indicator, as we designed a DM, we also compared each of the six objective functions used as performance metrics in the test oracles for the solutions proposed by the DM after the search budget was exceeded. 



\subsubsection{Statistical tests}

Since the employed algorithms are non-deterministic, we run each algorithm 10 times. We could not afford more runs given that the search budget was selected to be 12 hours. Therefore, in total we employed 10 (runs) $\times$ 12 (hours) $\times$ 2 (baseline and repair algorithms) = 240 hours for executing the experiments.

To assess the statistical significance, we employed the Wilcoxon rank sum test. We considered that there was statistical significance between both algorithms when the p-value was lower than 0.05. In addition, we employed the Vargha and Delaney \^{A}$_{12}$ value, which measures the probability of a technique being better than the other one.

\subsubsection{Algorithm configuration}

We gave 12 hours of time budget to both, our algorithm and the baseline algorithm. Similar to \cite{abdessalem2020automated}, the maximum number of solutions in the archive was set to 12 (i.e., 6 objective function $\times$ 2). We also set the parameter $N_{susp} = 5$, which means that the suspiciousness of a parameter is neutral (i.e., suspiciousness score of 0.5) until it is mutated 5 times.


\subsection{Analysis and Discussion of the Results}

\subsubsection{RQ1 -- Sanity check}

\begin{figure}[h]
\centering
\includegraphics[trim = 0 0 0 0, width=0.35\textwidth]{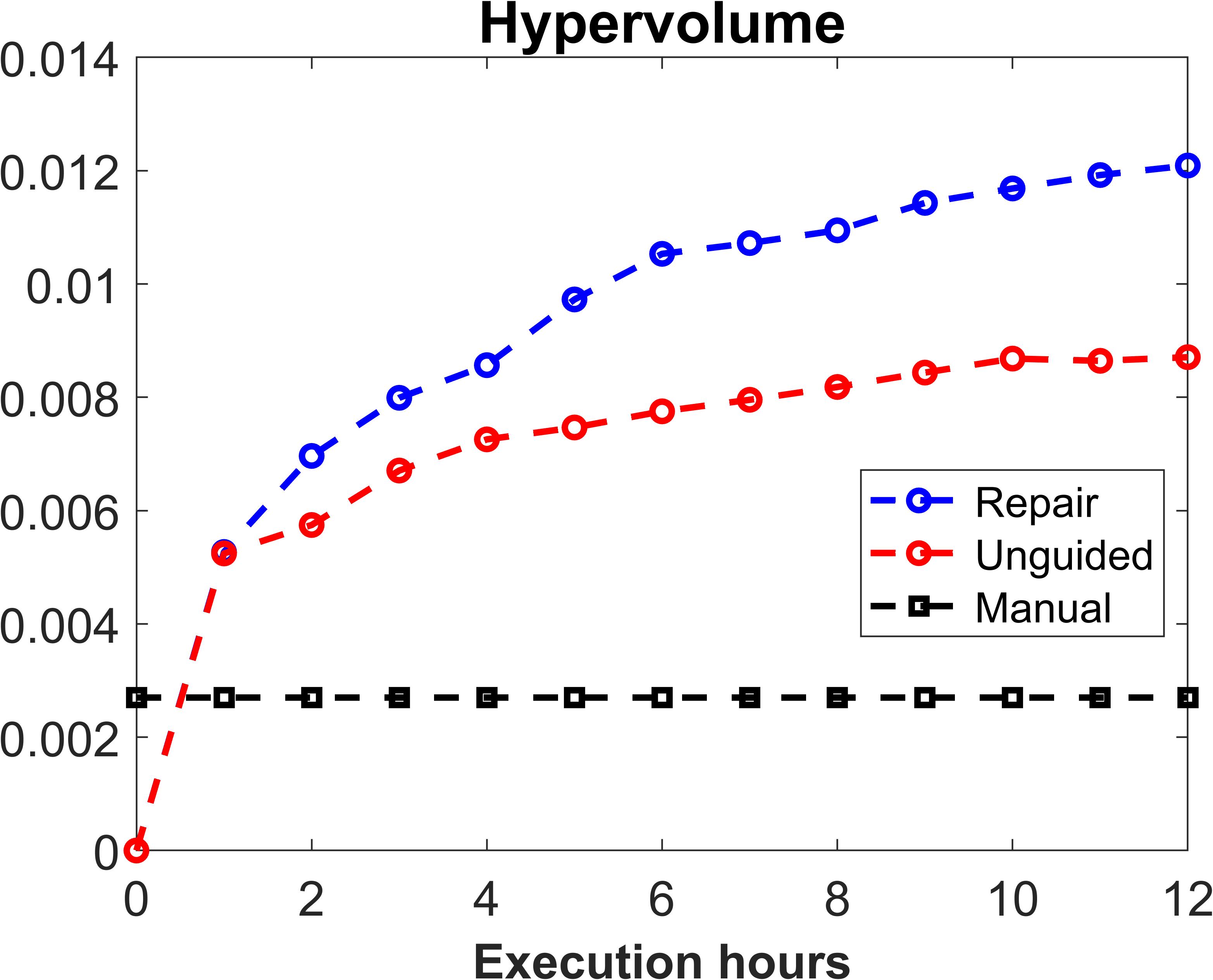} 
\caption{Average value of the 10 runs for the hypervolume quality indicator when comparing the repair and unguided algorithms}
\label{fig:results_HV}
\end{figure}

Figure \ref{fig:results_HV} shows the average \textit{HV} score of the 10 runs for both, the repair algorithm proposed in this paper and the baseline algorithm, which is the unguided version of the repair algorithm. As it can be appreciated, the repair approach showed a higher average \textit{HV} than the baseline after the second execution hour. By the time the search budget was expired, the repair algorithm showed a 29\% average improvement over the baseline in terms of the \textit{HV} quality indicator. It is noteworthy that the \textit{HV} values are quite low. The reasons for this is that the \textit{HV} favors knee points of a solution set in a Pareto-frontier~\cite{li2022evaluate}. As explained before, in our case, the specified threshold values were 0 (i.e., the repair algorithm aims at optimizing as much as possible all the functional performance metrics). Achieving such value was not realistic, and therefore we did not have knee values. Besides, 6 different oracles (i.e., fitness functions) were employed to guide the search towards providing patches. Nevertheless, a low \textit{HV} value makes not unfair the comparison between both techniques, which is the goal of the first RQ.

These results were further corroborated by means of statistical tests. Table \ref{tab:RQ1_statisticaltests} shows the \^{A}$_{12}$ as well as p-values (computed by employing the Wilcoxon rank sum test) for each of the 12 hours when comparing the repair algorithm with the baseline. The \^{A}$_{12}$ shows the probability of the repair algorithm being better than its unguided version. As suggested by Romano et al.~\cite{romano2006exploring}, we categorized the difference existing between the repair algorithm and its baseline as \textit{negligible} if $d<0.147$, as \textit{small} if $d<0.33$, as \textit{medium} if $d<0.474$ and as \textit{large} if $d>=0.474$, where $d=2|$\^{A}$_{12}-0.5|$. According to this categorization, the difference was negligible during the first execution hour, small between the second and third execution hours and medium during the fourth execution hour. In these first four execution hours, there was no statistical significance between the repair algorithm and the baseline. Conversely, after the fifth hour, there was statistical significance (i.e., p-value $<$  0.05) with large effect sizes based on the related categorization~\cite{romano2006exploring}, all of them in favor of our approach.

\begin{table}[h]
\centering
\caption{RQ1 -- Summary of the statistical tests when comparing the repair algorithm with its unguided version for the \textit{HV} quality indicator over the execution of 12 hours. An \^{A}$_{12}$ value higher than 0.5 means that the results are in favor of the repair algorithm. Statistical significance is set as~\textit{p-val$<$0.05}}
\begin{tabular}{rrr}
\hline
\multicolumn{1}{l}{\textbf{Hour}} & \multicolumn{1}{l}{\textbf{\^{A}$_{12}$}} & \multicolumn{1}{l}{\textbf{p-val}} \\ \hline
1                                   & 0.51                              & 0.9698                                \\ 
2                                   & 0.61                              & 0.4273                                \\ 
3                                   & 0.65                              & 0.2730                                \\ 
4                                   & 0.71                              & 0.1212                                \\ 
5                                   & 0.80                               & 0.0256                                \\ 
6                                   & 0.86                              & 0.0081                                \\ 
7                                   & 0.89                              & 0.0040                                \\ 
8                                   & 0.82                              & 0.0172                                \\ 
9                                   & 0.85                              & 0.0090                                \\ 
10                                  & 0.85                              & 0.0090                                \\ 
11                                  & 0.90                               & 0.0028                                \\ 
12                                  & 0.92                              & 0.0017                                \\ \hline
\end{tabular}
\label{tab:RQ1_statisticaltests}
\end{table}

Besides the \textit{HV}, we also analyzed the individual patches provided by the decision maker (DM). In this case, the aim of the algorithm was to reduce such metrics. Therefore, an \^{A}$_{12}$ lower than 0.5 means that the repair algorithm performed better. Table \ref{tab:Indiv_objs2} summarizes the statistical tests for the ten runs and each individual objective function. There was statistical significance in half of the objective functions (i.e., LWT, ATT and LTT). For such cases, the effect sizes were large (i.e., \^{A}$_{12}$ between 0.18 to 0.2). For the remaining objectives, where there was no statistical significance, in the case of the AWT and \%WT$>$55, the effect sizes showed a negligible difference, whereas for the case of \%TT$>$70, the difference was small.

\begin{table}[h]

\centering
\caption{Summary of the statistical test results when comparing the patches provided by the DM when employing repair algorithm against the baseline and manual repair approaches}

\begin{tabular}{lrrrr}
\cline{2-5}
                                            & \multicolumn{2}{l}{\textbf{vs. Baseline}}                      & \multicolumn{2}{l}{\textbf{vs. Manual}}                        \\ \cline{2-5} 
                                            & \multicolumn{1}{l}{\textbf{\^{A}$_{12}$}}  & \multicolumn{1}{l}{\textbf{p-val}} & \multicolumn{1}{l}{\textbf{\^{A}$_{12}$}}  & \multicolumn{1}{l}{\textbf{p-val}} \\ \hline
\multicolumn{1}{l}{AWT}                   & \multicolumn{1}{r}{0.52} & 0.9097                     & \multicolumn{1}{r}{0.10} & 0.0014                     \\ 
\multicolumn{1}{l}{LWT}                   & \multicolumn{1}{r}{0.18} & 0.0165                     & \multicolumn{1}{r}{0.20} & 0.0161                     \\ 
\multicolumn{1}{l}{\%WT\textgreater{}55s} & \multicolumn{1}{r}{0.47} & 0.8788                     & \multicolumn{1}{r}{0.00} & \textless{}0.0001          \\ 
\multicolumn{1}{l}{ATT}                   & \multicolumn{1}{r}{0.20} & 0.0312                     & \multicolumn{1}{r}{0.40} & 0.4429                     \\ 
\multicolumn{1}{l}{LTT}                   & \multicolumn{1}{r}{0.20} & 0.010                      & \multicolumn{1}{r}{0.00} & \textless{}0.0001          \\ 
\multicolumn{1}{l}{\%TT\textgreater{}70s} & \multicolumn{1}{r}{0.37} & 0.3438                     & \multicolumn{1}{r}{0.00} & \textless{}0.0001          \\ \hline
\end{tabular}
\label{tab:Indiv_objs2}

\end{table}




Table \ref{tab:averages} show the average value of each of the functional performance metrics used by the oracles for the 10 runs and the patches provided by the DMs. These results were somehow consistent with those from Table \ref{tab:Indiv_objs2}. As it can be appreciated, the most striking difference relates to the LWT and the LTT functional performance metrics. On the contrary, for the AWT, \%WT$>$55, ATT and \%TT$>$70, the differences were not that large. This could be due to the nature of the DM. Note that for those metrics, the DM accepts values that are below certain thresholds (e.g., AWT $<$ 25 seconds), whereas for LWT and LTT, the DM selects those patches with lowest values. However, in all metrics except the AWT, our algorithm showed lower average values.

\begin{table}[ht]
\caption{Comparison between the misconfigured configuration file, the patch provided by the DM with the manual repair, the average values of the patches returned by the DM for the baseline algorithm and the average values of the patches returned by the DM for the repair algorithm}
\begin{tabular}{lrrrr}
\cline{2-5}
 & \multicolumn{1}{c}{\textbf{Misconf}} & \multicolumn{1}{l}{\textbf{Manual}} & \multicolumn{1}{l}{\textbf{Baseline DM}} & \multicolumn{1}{l}{\textbf{Repair DM}} \\ \hline
AWT & 25.99 & 23.10 & 22.66 & 22.77 \\
LWT & 435.70 & 223.00 & 241.55 & 213.72 \\
\%WT \textgreater{}55s & 12.78 & 11.99 & 9.93 & 9.92 \\
ATT & 42.01 & 41.60 & 41.77 & 41.58 \\
LTT & 209.80 & 220.60 & 206.24 & 195.56 \\
\%TT\textgreater{}70s & 10.24 & 10.02 & 9.64 & 9.45 \\ \hline
\end{tabular}
\label{tab:averages}
\end{table}

In conclusion, the first RQ can be answered as follows:

\vspace{0.25cm}
\fbox{\begin{minipage}{23 em}
\textbf{Answer to the first RQ: }\textit{The repair algorithm outperformed the baseline algorithm. The average improvement extent of the repair algorithm with respect to the baseline was around 29\% when considering the \textit{HV} quality indicator. Furthermore, there was statistical significance with large effect sizes when comparing individual patches proposed by the DM for half of the objective functions, all of them in favor of the repair algorithm. All this suggests that the problem of repairing CPSs misconfigurations is non-trivial, and therefore, automated and scalable repair techniques are necessary.}
\end{minipage}}
\vspace{0.25cm}

\subsubsection{RQ2 -- Comparison with manual repair}

With the second RQ, we aimed at comparing the proposed repairing algorithm with the manual process of repairing the misconfiguration by domain experts. Specifically, these domain experts provided a total of 6 patches. With those patches, and by applying the six oracles in our algorithm, we derived the \textit{HV} metric. As can be seen in Figure \ref{fig:results_HV}, the \textit{HV} was quite low. This was because only four patches were non-dominated, whereas our archive is capable of handling up to twelve patches. Therefore, those four patches were not able to cover a large volume in the objective space. Furthermore, it is important to note that the time was not considered here, because we do not have such information. In terms of the HV, the average improvement extent of our repair algorithm over the manually derived patches was up to 77.5\%.

For this case, we also employed the DM to select one of the non-dominated patches. Table \ref{tab:Indiv_objs2} shows the statistical tests carried out when comparing the patches provided by the DM after executing the repair algorithm with the patch proposed by the DM after processing the four non-dominated solutions. As it can be appreciated, in five out of six metrics there was statistical significance, where the effect size showed a large difference according to the categorization proposed by Romano et al.~\cite{romano2006exploring}. All these effect sizes were in favor of the repair algorithm. On the other hand, for the case where there was no statistical significance, i.e., for the case of the ATT metric, the difference was small in terms of the \^{A}$_{12}$ value, but in favor of the repair algorithm.

The improvement extent for each functional performance metric obtained by the patches provided by the DM (over 10 runs) with respect to the manual approach can be appreciated in Table \ref{tab:averages}. These results are consistent with the statistical tests, where it can be appreciated a similar average value in the case of the ATT. In this case, the improvement extent is higher in the cases of the AWT, \% WT $>$ 55, LTT and \%WT$>$70 when compared to the baseline algorithm. However, in relation to the LWT, the improvement was only of 10 seconds on average, unlike with the baseline, where the improvement was of nearly 29 seconds on average.

In summary, the second RQ can be answered as follows:

\vspace{0.25cm}
\fbox{\begin{minipage}{23 em}
\textbf{Answer to the second RQ: }\textit{The repair algorithm outperformed the manual repair process. The average improvement extent of the repair algorithm with respect to the patches provided by the domain experts was around 77.5\% when considering the \textit{HV} quality indicator. Furthermore, there was statistical significance with large effect sizes when comparing individual patches proposed by the DM in five out of six objective functions. In addition, our approach provides a fully automated approach, which can therefore increase the productivity of engineers from Orona when dealing with misconfigurations of the traffic dispatching algorithm.}
\end{minipage}}
\vspace{0.25cm}








\subsection{Threats to Validity}


We now summarize the threats to validity of our study and the measures taken to mitigate them. 

An \textit{internal validity threat} in our evaluation could be related to the parameters used in the algorithms, which were not changed. Three main parameters need to be configured (1) the time budget, which was set to 12 hours; (2) the number of time a parameter needs to be selected to start computing its suspiciousness score (i.e., $N_{susp}$), which is set to 5; and (3) the number of solutions in the archive. The first two parameters were selected based on preliminary evaluations. Coversely, the maximum number of solutions in the archive was the same as other repair approaches targeting CPSs~\cite{abdessalem2020automated}. 

As in any search-based software engineering problem, a \textit{conclusion validity threat} involves the stochastic nature of the algorithms used. To mitigate such issue, we run each algorithm 10 times. It is important to note that our technique needs a long time to converge because the simulations employed to assess potential patches are exhaustive, therefore, we could not afford a large number of runs. Furthermore, we applied statistical tests to analyze the results, as recommended by Arcuri and Briand~\cite{arcuri2011practical}. 

As in any study involving humans, our evaluation is also subject to \textit{external validity threats}. One such threats refers to the patches proposed by engineers from Orona. It is noteworthy, however, that these engineers have broad experience and domain expertise, and that the patches they proposed were the ones that were later deployed in the real CPS. The generalizability of the results is also another \textit{external validity threat} of our study; note, however, that we used an industrial case study with a real installation and data obtained from operation. We plan to mitigate such threat in the future by (1) using other case studies from a different domain and (2) using other real installations where misconfigurations occured.

Lastly, \textit{construct validity threats} arise when the measures used are not comparable across algorithms. This was mitigated by giving the same search budget to both algorithms (i.e., the repair and the unguided algorithm).

\section{Lessons Learned and Applicability}
\label{sec:lessons}
In this section, we describe the lessons we have learned thorough the whole process of developing and evaluating the repairing algorithm. In addition, we explain the main changes our method would require when applying it to other CPS domains.

\subsection{Lessons Learned}

\textbf{Lesson 1 -- Reduction of personnel cost:} The current state of the practice when repairing misconfigurations is purely manual. This requires significant personnel cost since domain experts are required in the process. Our fully automated repairing approach not only outperforms the state of the practice in terms of providing a better patch to repair the misconfiguration, but also reduces significantly the personnel costs that are required behind a manual repair process.

\textbf{Lesson 2 -- Scalable technique:} Scalability is one of the main concerns when testing and debugging CPSs, mainly due to the need of considering properties involving physical devices with continuous dynamics and complex concurrent interactions between the system and its environment (e.g., people)~\cite{briand2016testing}. We saw that our search-based repair algorithm converges after around 10 hours, which is affordable for our industrial partner as engineers can launch the automated misconfiguration repair tool nightly.

\textbf{Lesson 3 -- Surrogate models are, in principle, not appropriate:} Despite we did not carefully assess this, while we developed the algorithm, we intended to integrate surrogate models to accelerate the repair process. However, we saw that this technique required too much time to build reliable surrogate models. This time was similar to the time budget that our repair algorithm required to converge. Although we assessed different types of surrogate models, we still need to more carefully analyze this, which remains a future work.

\textbf{Lesson 4 -- Challenging conflicting installation:} After applying our experiments and showing the results to Orona's engineers, we noted that the conflicting installation we selected was challenging. Indeed, the traffic was abnormal, with many unforeseen situations (e.g., having too many calls in a short time window) and therefore, repairing the misconfiguration in such installation was, according to domain experts, more difficult than other installations.


\subsection{Applicability}

The context at which we have applied our repairing approach is the elevator dispatching algorithm of Orona. However, we believe that the three key challenges that we tackle (i.e., expensive execution of tests, large configuration space and multiple functional performance requirements) are common in all types of configurable CPSs. As we involved domain experts when developing the repair approach, several domain-specific design choices were considered, which would require adaptions when applying our approach in another domain. Below we explain different alternatives and the changes required for the adoption of our method in another domain.

\textit{\textbf{Test execution process: }}One of the first changes our method would require is the test execution. As explained in Section \ref{sec:testexec}, we use a domain-specific simulator to execute test cases and measure how close the algorithm is from repairing the misconfiguration. This process would need to be substituted by the simulator being used to execute the tests within other CPSs. In addition, we employ a parallel test execution, which was possible in our context. However, other simulators (e.g., autonomous vehicles) could require more computing resources. For instance, testing autonomous vehicles often requires rendering driving scenes in virtual scenarios using high-fidelity simulators~\cite{haq2022efficient}, which may require the execution of test cases to be sequential. Lastly, test oracles would need to be defined. When using Simulink models to execute the tests, which is a predominant CPS testing tool~\cite{matinnejad2018test}, an option could be to use SOCRaTEs~\cite{menghi2019generating}, a DSL-based test oracle specification and generation tool for Simulink. Specifically, SOCRaTEs~\cite{menghi2019generating} provides a quantitative measure of the degree of violation of a requirement, similar to what we need in our algorithm to guide the misconfiguration repair process.


\textit{\textbf{Removing solutions from the archive: }}As explained in Section \ref{sec:archive}, the archive may increase in size, which may have a direct implication in the convergence of the repairing algorithm. Therefore, when the archive exceeds a predefined number of solutions, one of the solutions needs to be removed. Our algorithm removes the solution with longest AWT, given that this is the most widely employed metric when testing dispatching algorithms~\cite{barney2015elevator}. In another domain, two alternatives can be considered. The first one, employing one of the most important metrics. If all metrics have a similar importance, the second alternative could be to randomly remove one of the solutions from the archive or use a crowding distance to remove solutions that are too close from each other.

\textit{\textbf{Decision maker: }}The decision maker is another component that we developed ad-hoc for the traffic dispatching algorithm by following the advise of domain experts. We recommend to analyze priorities of the specific CPS to make a decision. In case there are no clear priorities, a solution could be to employ a weighted approach giving the same importance to all objectives.

\textit{\textbf{Patch confirmation: }}We only employed a failing test suite to guide the repair process. The core reason was the high test execution time. Eventually, it could happen that a proposed patch makes a test case from the passing test suite fail. Because of this, we implemented a patch confirmation process by following a traditionally employed regression test method~\cite{gartziandiamachine} combined with a newly incorporated metamorphic testing approach by Orona~\cite{ayerdi2020qos,ayerdi2022performance}. The patch confirmation module should follow the internally standardized testing approach, which can vary from a company to another.

\section{Related Work}\label{sec:relatedwork}

The related work in automated program repair is huge. Monperrus mantains a living review on such techniques~\cite{monperrus2018living}. Table~\ref{tab:RelatedWork} shows a summarized classification of the related work analyzing four key characteristics covered by our approach. The first characteristic (\textbf{C1}) analyzes the possibility of repairing computationally expensive systems. The second one (\textbf{C2}), whether the approach is intended to repair misconfigurations. The third one (\textbf{C3}), analyzes if the approach is able to deal with many requirements (i.e., more than 3). And the last one (\textbf{C4}), whether the approach prioritizes critical faults over the less critical ones.

\begin{table}[h]
\centering
\caption{Related work comparison with different characteristics required by our repairing technique}
\label{tab:RelatedWork}
\begin{tabular}{lcccc}
\hline
                              & \multicolumn{1}{l}{\textbf{C1}} & \multicolumn{1}{l}{\textbf{C2}} & \multicolumn{1}{l}{\textbf{C3}} & \multicolumn{1}{l}{\textbf{C4}} \\ \hline
\cite{abdessalem2020automated}                         & +                                & -                                & +                                & +                                \\ 
\cite{krishna2020cadet,xiong2014range}                         & -                                & +                                & -                                & -                                \\ 
\cite{jung2021swarmbug}                   & +                                & +                                & -                                & -                                \\ 
\cite{ackling2011evolving,kim2013automatic,knowles2001reducing,le2011genprog,gissurarson2022propr}                & -                                & -                                & -                                & -                                \\ 
\cite{arcuri2008automation,dallmeier2009generating,demarco2014automatic,ji2016automated,nguyen2013semfix, qi2014strength, qi2015analysis,weimer2013leveraging}  & +                                & -                                & -                                & -                                \\ \hline
\end{tabular}
\end{table}

We found that, in the field of CPSs, repairing approaches are still in their infancy. Indeed, to the best of our knowledge, only two approaches tackle the problem of repairing CPSs. On the one hand, Swarmbug~\cite{jung2021swarmbug} focuses on repairing misconfigurations of swarm robotics. Specifically, they make use of a mechanism called the ``\textit{degree of causal contribution}'' to abstract impacts of configurations to the swarm drones via behavior causal analysis. The evaluation is carried out in four swarm algorithms, and the repair objectives are individual for each of them. These involve aspects like avoiding obstacles or unsafe zones in order the drones not to crash. The approach, however, does not cover \textbf{C3} and \textbf{C4}. On the other hand, Ariel~\cite{abdessalem2020automated} focuses on repairing feature interaction failures in automated driving systems. Similar to our approach, ARIEL~\cite{abdessalem2020automated} uses a many-objective and a single population-based approach, and also employs an archive to keep track of partially repaired solutions. However, unlike this paper, which focuses on repairing misconfigurations, ARIEL~\cite{abdessalem2020automated} repairs feature interaction bugs by applying \textit{modify} and \textit{swift} mutation operators to integration rules that resolve conflicts between automated driving system features. Therefore, ARIEL does not cover \textbf{C2}.


CADET~\cite{krishna2020cadet} does cover \textbf{C2} as it is intended to debug and fix misconfigurations that cause non-functional faults. Xiong et al.~\cite{xiong2014range} focus on repairing misconfigurations in software product lines by generating a list of range fixes to help satisfy a constraint. However, both approaches do not consider systems that take high computation resources to execute the tests. In addition, CADET~\cite{krishna2020cadet} only covers two non-functional properties (i.e., latency and energy), whereas Xiong et al.~\cite{xiong2014range} focus on satisfying individual constraints. Lastly, the approaches do not prioritize fixing more critical faults over the less critical ones. Subsequently, both techniques do not cover \textbf{C1}, \textbf{C3} and \textbf{C4}.

Besides these three studies, which are the most closely related to our approach, other studies exist in the field of automated program repair~\cite{ackling2011evolving,kim2013automatic,knowles2001reducing,le2011genprog, gissurarson2022propr, arcuri2008automation,dallmeier2009generating,demarco2014automatic,ji2016automated,nguyen2013semfix, qi2014strength, qi2015analysis,weimer2013leveraging}. Similar to this approach, some consider search techniques, such as genetic programming~\cite{le2011genprog,gissurarson2022propr}. GenProg~\cite{le2011genprog} is one of the first approaches that proposed the use of meta-heuristic search to repair software programs. Specifically they leveraged genetic programming to repair C programs. However, all these approaches focus on repairing bugs in the code. Conversely, our approach focuses on repairing misconfigurations in the field of configurable CPSs. 

Another line of research related to our approach is that of unified debugging~\cite{lou2020can,benton2021evaluating}. Such technique uses patch execution results to improve localizing the fault~\cite{lou2020can,benton2021evaluating}. Therefore, even if the repair process is unable to repair the bug, unified debugging helps improving the fault localization for latter manual repair. Our approach follows a similar strategy, where we aim at localizing suspicious parameters that will eventually help repair the misconfiguration. However, besides the fact that unified debugging~\cite{lou2020can,benton2021evaluating} is not aimed at debugging misconfigurations, but bugs at the code level, it assumes that there is an initial suspiciousness score (i.e., at statement level). Conversely, our approach begins with all parameters having the same suspiciousness because there is no information about which parameters have influence in the system performance.

\section{Conclusion and Future Work}
\label{sec:conclusion}

Real-world CPSs, such as elevators, involve many parameters. The performance of CPSs is tightly linked to such parameters, and therefore, misconfigurations may occur. On the one hand, manually dealing with such misconfigurations might not always be feasible. On the other hand, automated solutions require dealing with certain challenges, such as, expensive simulations to execute test cases. In this paper we propose an automated and scalable solution based on meta-heuristic search to repair misconfigurations in CPSs. Our approach was integrated with an industrial case study provided by Orona, one of the largest elevator manufacturers in Europe. The evaluation was carried out with a real installation in which domain experts from Orona had to manually intervene in repairing a misconfiguration. The results suggest that, besides automating a process that before was purely manual, our algorithm provides better patches than those provided by domain experts. Specifically, in five out of the six quality indicators employed by domain experts to assess the quality of a patch, our algorithm outperformed with statistical significance the patch provided by domain experts.

In the future, we would like to extend our approach from different perspectives. In terms of the applicability, we would like to integrate our algorithm with other CPSs in which configurations have been found to be problematic (e.g., unmanned aerial vehicles~\cite{han2022control}). Furthermore, we would like to explore solutions to prevent potential overfitting issues before proposing a plausible patch. This has been one of the core challenges identified in automated program repair~\cite{goues2019automated,martinez2017automatic,nilizadeh2021exploring,smith2015cure}, and therefore, we should be aware of it. In terms of internal applicability within Orona, we would like to evaluate our approach in other installations where misconfigurations occurred. Furthermore, we would also like to transfer the repair algorithm beyond the traffic team and within other departments. Lastly, we would like to further study whether other strategies exist to better train and integrate surrogate models in the repair process.



%



\ifCLASSOPTIONcompsoc
  \section*{Acknowledgments}

\else
  \section*{Acknowledgment}
\fi

 Project supported by a 2021 Leonardo Grant for Researchers and Cultural Creators, BBVA Foundation. The BBVA Foundation is not responsible for the opinions, comments and contents included in the project and/or the results derived from it, which are the total and absolute responsibility of their authors. Aitor Arrieta is part of the Software and Systems Engineering research group of Mondragon Unibertsitatea (IT1519-22), supported by the Department of Education, Universities and Research of the Basque Country.

\bibliographystyle{IEEEtran}
\bibliography{biblography}

\ifCLASSOPTIONcaptionsoff
  \newpage
\fi

\end{document}